\documentclass[preprint,showkeys,longbibliography]{revtex4-1}

\usepackage[active]{srcltx}
\usepackage{graphicx}
\usepackage{amsmath}
\usepackage{mathcomp}	
\usepackage{hyperref}
\usepackage{color}

\begin{document}

\title{PtSi Clustering In Silicon Probed by Transport Spectroscopy}

\author{Massimo Mongillo}
\altaffiliation{Present address: CEA-Leti MINATEC Campus, 17 Rue des Martyrs, 38054 Grenoble, France, E-mail: massimo.mongillo@cea.fr}
\affiliation{SPSMS/LaTEQS, CEA-INAC/UJF-Grenoble 1, 17 Rue des Martyrs, 38054 Grenoble Cedex 9, France}

\author{Panayotis Spathis}
%\altaffiliation{Present address: Institut N\'eel, CNRS and Universit\'e Joseph Fourier, B.P. 166, 38042 Grenoble, France}
\affiliation{SPSMS/LaTEQS, CEA-INAC/UJF-Grenoble 1, 17 Rue des Martyrs, 38054 Grenoble Cedex 9, France}

\author{Georgios Katsaros}
%\altaffiliation{Present address Johannes Kepler University, Institute of Semiconductor and Solid State Physics, Altenbergerstrasse 69, 4040 Linz, Austria}
\affiliation{SPSMS/LaTEQS, CEA-INAC/UJF-Grenoble 1, 17 Rue des Martyrs, 38054 Grenoble Cedex 9, France}

\author{Riccardo Rurali}
\affiliation{Institut de Ci\`{e}ncia de Materials
             de Barcelona (CSIC), Campus de Bellaterra,
             08193 Bellaterra, Spain}
             
\author{Xavier Cartoix\`{a}}
\affiliation{Departament d'Enginyeria Electr\`{o}nica,
             Universitat Aut\`{o}noma de Barcelona,
             08193 Bellaterra, Spain}
            
\author{Pascal Gentile}
\affiliation{SP2M/SINAPS, CEA-INAC/UJF-Grenoble 1, 17 Rue des Martyrs, 38054 Grenoble Cedex 9, France}
%\author{Marc Sanquer}
%\affiliation{SPSMS/LaTEQS, CEA-INAC/UJF-Grenoble 1, 17 Rue des Martyrs, 38054 Grenoble Cedex 9, France}
\author{Silvano De Franceschi}
\affiliation{SPSMS/LaTEQS, CEA-INAC/UJF-Grenoble 1, 17 Rue des Martyrs, 38054 Grenoble Cedex 9, France}
\email{silvano.defranceschi@cea.fr}

%\author{Massimo Mongillo}
%\email{massimo.mongillo@gmail.com}
%\affiliation{SPSMS/LaTEQS, }
%\homepage{http://legauss.blogspot.com}
%\author{Georgios Katsaros}
%\author{Panayotis Spathis}
%\email{johndoe@someserver.com}
%\affiliation{SPSMS/LaTEQS, CEA-INAC/UJF-Grenoble 1, 17 Rue des Martyrs, 38054 Grenoble Cedex 9, France}

%\author{W. Y. S. I. W. Y. Get}
%\email{wisiwig@duh.com}
%\homepage{http://www.google.com}
%\affiliation{National Other Institute of Other Country.}

\date{\today}

\begin{abstract}
Metal silicides formed by means of thermal annealing processes are employed as contact materials in microelectronics. Control of the structure of silicide/silicon interfaces becomes a critical issue when the device characteristic size is reduced below a few tens of nanometers. Here we report on silicide clustering occurring within the channel of PtSi/Si/PtSi Schottky barrier transistors. This phenomenon is investigated through atomistic simulations and low-temperature resonant tunneling spectroscopy. Our results provide evidence for the segregation of a PtSi cluster with a diameter of a few nanometers from the silicide contact. The cluster acts as metallic quantum dot giving rise to distinct signatures of quantum transport through its discrete energy states.  
\end{abstract}


\maketitle

%\begin{abstract}

In the recent years extensive research has been devoted to the replacement of the heavily-doped silicon contact regions of conventional metal-oxide-semiconductor fied-effect-transistors (MOSFETs) with metallic binary compounds of silicon and a transition element, generally referred to as metal silicides \cite{Larson2006}. This approach aims at reducing source/drain contact resistances and, simultaneously, relaxing the constrains imposed by doping-profile control. In addition, a lower thermal budget can be expected from the use of relatively low-temperature silicidation processes, as opposed to the high-temperature annealing cycles required for dopant activation.
The removal of heavily doped contact regions is not priceless though. In fact, an undesirable Schottky barrier (SB) is introduced with a negative impact on device performances.  The SB reduces the on-state current and decreases the sub-threshold slope, leading to a poorer switching performance of the MOSFET. This problem can be mitigated by choosing silicides with low SB height. The high working function of platinum silicide (PtSi) makes this compound the best candidate for contacts to p-type silicon channels. The reported values for the SB height, $\phi_{B}$, in PtSi/p-Si junctions range between $0.15$ and $0.27  $eV \cite{Dubois2004}. 
The attractiveness of PtSi is not only limited to p-channel SB transistors. This metal silicide has been successfully used also for the fabrication of inverters in complementary MOS technology where it forms the contact material of both p-type and n-type transistors \cite{Larrieu2011}. In the latter case, the formation of the silicide is accompanied by the accumulation of n-type dopants (typically As or P) close to the PtSi/n-Si interface leading to a suppression of the effective SB height. This approach avoids the integration of a second silicide for n-type contacts,  typically ErSi\cite{Jang2005} or YSi\cite{Zhu2004}. Furthermore, PtSi is routinely used in p-type Schottky diodes for infrared photo detection \cite{Shigeyuki1992, Lin1995}. 

PtSi, and metal silicides in general, are formed through a thermally activated process. Upon annealing of a thin Pt film on Si,  the two materials diffuse into each other \cite{Poate1974}. (This differs from the case of other commonly used silicides, e.g. nickel silicide, where essentially only the metal element acts as a diffusing species.) Contact fabrication relies on an accurate control of 
this inter-diffusion process, which leads to the formation of a silicide phase. In particular, the achievement of sharp boundary between PtSi and Si is important in order to meet the demand for device scaling down to characteristic channel lengths of only a few tens of nm.  Non-abrupt interfaces, and the possible simultaneous diffusion of metal impurities into the channel region, can alter important figures of merit such as the sub-threshold slope of the on-state current leading to an unacceptably high device variability \cite{Asenov2003}. 
% ggvwith a consequent fluctuation of the threshold \cite{PierreM.2010}, SS and SB height \cite{Calvet2002} from device to device.
Although thermal annealing can result in atomically abrupt silicide/silicon interfaces \cite{Lin2008}, often this is not the case and various degrees of interface roughness can be found\cite{Dellas2012-09-01,Laszcz2006,Liu2007}. In addition, several experimental evidences of unintentional Pt impurities in the channel of silicon MOSFETs have been reported \cite{Calvet2008,Calvet2011,Calvet2008b}. Such Pt atoms originate during the silicidation process as a result of a diffusive motion from the PtSi/Silicon interface into the Si channel. Yet little is known about their most favorable arrangement within the silicon crystal and, in particular, about how this arrangement is affected by collective interactions among multiple Pt impurities. 
%Answering this question is crucial in order to understand the impact of Pt impurities and to evaluate their effect on the device electrical properties. 
In this Letter we address this problem through a combined theoretical and experimental study involving atomistic simulations and transport measurements in short-channel PtSi/Si/PtSi SB transistors. 

%They give rise to signatures of resonant tunnelling \cite{Fowler1986} through impurity states located in the Silicon band-gap. 

% \fcolorbox{red}{yellow}{HERE RESULTS FROM SIMULATIONS:}

%Silicidation of Platinum with Silicon can lead to the nucleation of PtSi clusters in the Silicon matrix. 

%In this letter we show it by atomistic simulations and we report experimental evidence of transport through an individual PtSi cluster.

%The thermal energy provided during a silicidation process promotes the formation of a PtSi phase through the inter-diffusion of Si atoms. 

We begin from the simplest case of a single Pt impurity in a Si lattice. Some experimental studies suggest that isolated Pt impurities can indeed be found in the Si region adjacent to a PtSi/Si interface~\cite{Mantovani1986,Prabhakar1983}. It has already been shown experimentally and theoretically  that in such a case the substitutional position (i.e. a Pt impurity at the place of a Si atom) is the most energetically favorable configuration~\cite{Watkins1995,Anderson1991,Anderson1992,Woodbury1962}. Here we consider the problem of multiple Pt impurities in a Si lattice. We intend to evaluate the possibility that nearby impurities out-diffusing from a PtSi contact can aggregate into small clusters. We also intend to find the most stable cluster structure.  

In order to tackle the problem of multiple Pt impurities we used numerical calculations based on density-functional theory~\cite{dft}. Starting from a single substitutional Pt, 
we identified the most energetically favorable position for a second substitutional Pt. This identification was accomplished by comparing three different options: first, second, and third neighboring lattice site. As a next step, starting from the most stable two-atom configuration, we studied three possible scenarios for the addition of a third Pt atom.
Notice that, given the large difference between the formation energies of interstitial and substitutional single Pt, we ruled out aggregates involving interstitial Pt.

\begin{figure}[h]
\includegraphics[width=8.5cm,keepaspectratio]{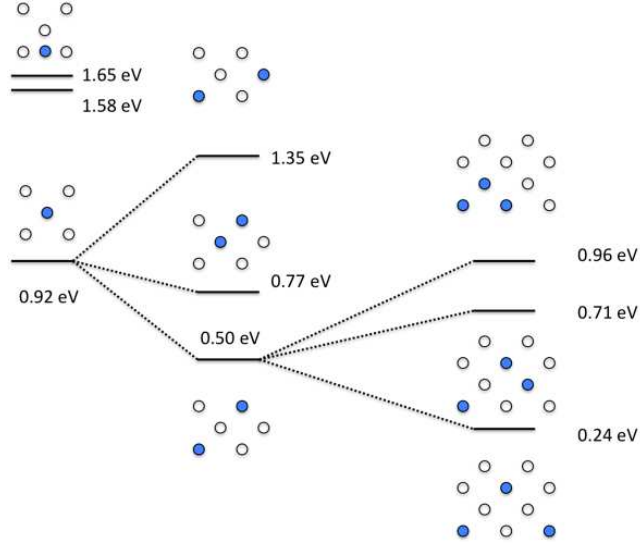}
\caption{Energy scales involved in the formation of a substitutional 
         Platinum defect in the silicon lattice. The most stable 
         configuration of a multi-atom defect is the one in which 
         Platinum atoms are second neighbors.
         Left column: formation energy of Pt point defects, i.e. 
         substitutional (0.92~eV) and interstitials (1.58 and 
         1.65~eV). Center column: formation energy per Pt atom of
         two-atom aggregates. First- and second-neighbors clusters
         favor aggregation, while a third-neighbor cluster does not.
         Right column: addition of a third Pt atom to the most stable
         of the two-atom clusters, i.e. the second-neighbor aggregate
         of the center column. The formation energy per impurity
         decreases only for the all-second-neigbour cluster (0.24~eV
         against 0.5~eV of the two-atom aggregate).
         }
\label{fig1}
\end{figure}

A branched diagram summarizing the results of these total-energy 
calculations is shown in Fig.~\ref{fig1}. Two observations 
can be made: (i)~a driving force promoting aggregation exists 
each time the formation energy per atom decreases following the 
addition of Pt atom; (ii)~given a set of possible configurations 
promoting aggregation, substitution as a second neighbor of the 
pre-existing Pt atom(s) is always preferred.
Hence, both first- and second-neighbor substitution are favored 
for the two-impurity aggregate, as the formation energy per atom 
decreases from 0.92~eV to 0.77 and 0.5~eV, respectively,
the latter being the most stable. When a third Pt atom reaches 
the second-neighbour two-atom aggregate, on the other hand, the 
only configuration that leads to an increased stability is the one 
where all the Pt atom are second neighbors. The other cases considered 
feature an increase of the formation energy per impurity (0.71 and 
0.96~eV)~\footnote{For the sake of simplicity we are assuming that, 
in the slow-rate limit, Pt atoms reaches the clustering zone one by 
one. We believe, however, that this simplified model is enough to 
capture the physics of the aggregation process.}.

\begin{figure}[h]
\includegraphics[width=8.5cm,keepaspectratio]{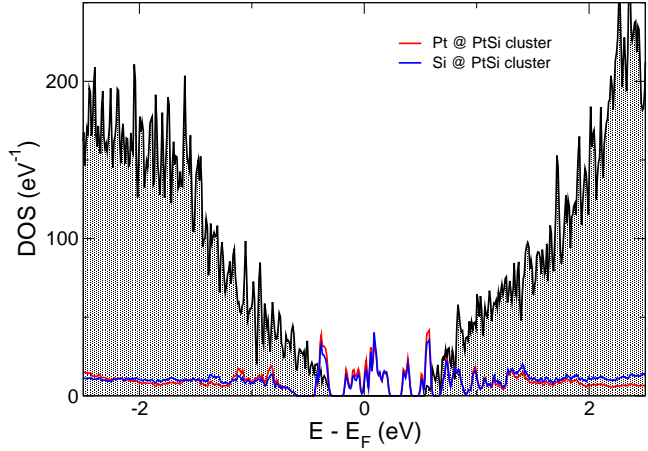}
\caption{ Total DOS for a PtSi cluster of 1.3 nm containing 32 Platinum atoms. The black shaded curve represents the DOS for bulk Silicon     }
\label{fig2}
\end{figure}

These results suggest that even a moderate supply of Pt atoms
into a Si lattice would lead to the formation of PtSi clusters
initially adopting the zincblende structure imposed by the host 
crystal. In order to test the validity of this conclusion for 
large numbers of Pt atoms, we considered a PtSi cluster with a 
diameter of approximately 1~nm, embedded in a 512-atom bulk Si supercell.
In Fig.~\ref{fig2} we plot the density of states decomposed in
the contributions from bulk Si atoms, and from Pt and Si atoms in
the PtSi cluster. It can be seen that a complex structure of
peaks appear in the Si band-gap, which reveals the metallic 
character of the cluster.

As the cluster grows in size, a phase transition to the bulk PtSi 
structure (favored by 0.55 eV/PtSi pair) is expected to occur. 
For this reason we have explicitly addressed the structural 
relaxation of a PtSi cluster of similar size with the thermodynamical stable structure, and we have found that for a cluster diameter of 1~nm 
relaxation to the host zincblende lattice is still thermodynamically 
favored.

We have also considered the formation of all-Pt clusters, examining 
both clusters where Pt adapts to the host zincblende symmetry or
clusters where Pt takes the fcc symmetry of its bulk form. Although 
the formation energies of Pt or PtSi inclusions depend on the chemical 
potentials of the respective constituent species (which have a large 
degree of uncertainty), the differences in favor of PtSi clusters 
are so large that all-Pt clusters can be safely discarded.

\begin{figure}[h]
\includegraphics[width=8.5cm,keepaspectratio]{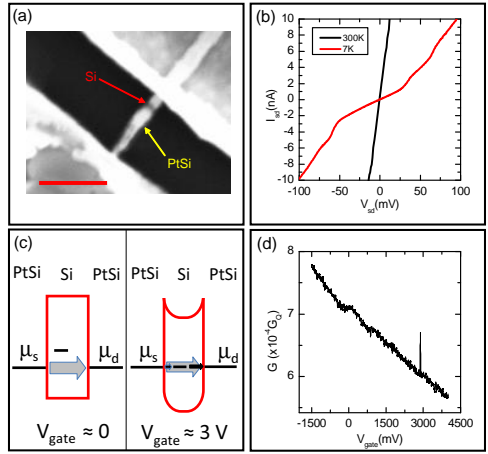}
\caption{(a)Scanning Electron Micrograph of the device before top gate patterning, showing the thin  semiconductor region (highlighted by the red arrow) sandwiched between the brighter platinum silicide contacts. The silicon channel between the two silicides is $12nm $ while the nanowire diameter is $40nm $. Scale bar 200nm.
(b) Current-voltage characteristics  taken at room temperature and $7$K of the device in (a).
(c) Simplified picture of the band diagram. At gate voltage $V_{gate} = 0$, where we assume flat-band condition in the silicon section, the cluster electrochemical potential lies above the Fermi levels of the contacts. In this regime, transport is  due to mainly hole-like direct tunneling through the silicon band gap. At $V_{gate} \approx 3 V $, the downward bending of the silicon bands results in a higher tunnel barrier for holes leading to a decrease in the direct-tunneling conductance. Simultaneously, the cluster electrochemical potential lines up with the Fermi levels of the leads resulting in a resonant-tunneling current.  
(d) A resonance peak appears over the background differential conductance  measured at $230mK $ with a lock-in excitation voltage of $100 \mu V $.  }
\label{fig3}
\end{figure}

In the following we shall present experimental evidence for the existence of PtSi clusters in silicon devices with PtSi contacts. We shall investigate the effect of a single PtSi cluster on the transport properties of a short-channel transistor.  
PtSi/Si/PtSi SB transistors were fabricated from undoped silicon nanowires (NWs) with diameters in the $20 - 40$ nm range. The NWs were grown by chemical vapor deposition via a catalytic vapor-liquid-solid method \cite{Wagner1964}.  After growth, the NWs were transferred onto an oxidized silicon substrate and individually contacted by pairs of 80-nm-thick Pt electrodes deposited by sputtering (for more details on NW growth and device fabrication we refer the reader to refs. \cite{Gentile2008,Mongillo2011}). 
Each Pt electrode consisted of a 500-nm-wide and 3-$\mu$m-long strip whose edges were connected to {\em two} Cr(10nm)/Au(65 nm) bonding pads via progressively wider Pt metal lines defined in the same deposition step. In order to promote the formation of PtSi contacts, the Pt strips were annealed one at a time by means of Joule effect. To this aim,  an electrical current of $\sim$10 mA was sequentially applied through each Pt strip causing a local increase of the temperature and hence promoting the silicidation of the contacts. This silicidation technique, which was introduced and extensively discussed in Ref. \cite{Mongillo2011}, was applied to obtain PtSi/Si/PtSi NW junctions with controlled Si channel length down to $\sim$10 nm (Fig.\ref{fig3}(a)). 
Each NW junction was capped by a 5-nm-thick aluminium-oxide (Al$_{2}$O$_{3}$) layer, grown by Atomic Layer Deposition, and a Cr(10nm)/Au(60nm) top-gate electrode defined by e-beam lithography, metal evaporation,  and lift-off.

Due to the absence of intentional doping and to the short channel length, the fabricated devices were found to operate as SB transistors \cite{Larson2006}. In these transistors, the silicon channel is fully depleted. The conduction- and valence-band edges have essentially flat spatial profile (see Fig. \ref{fig3}( c), left panel)  and their energy position relative to the Fermi levels of the PtSi contacts is set by the n- and p-type SBs, respectively. Since the p-type SB  ($\phi_p $) is significantly smaller than the n-type SB ($\phi_n$), electrical conduction  is dominated by hole-type carriers.. 
%At room temperature, most of the source-drain current is due to the thermionic emission of holes over the reversely biased p-type SB. Upon lowering temperature, this thermionic contribution decreases leaving the room to hole tunneling processes through the silicon section. 
Fig.\ref{fig3}(b) shows the source-drain current, $I_{sd}$, as a function of source-drain bias voltage, $V_{sd}$, for two different temperatures. At room temperature (black trace) and for $V_{sd}<<\phi_{B}$, transport through the silicon region is dominated by the thermionic emission of holes over the reverse-bias p-type SB. At 7 K (red trace), thermionic emission is entirely suppressed, and the residual conduction is due to temperature-independent tunneling through the silicon section, which acts as tunnel barrier \cite{Simmons1963}. 
A finite differential conductance, $G=dI_{sd}/dV_{sd}$, is observed throughout the entire  $I_{sd}(V_{sd})$, thus including the linear regime around zero bias. The linear conductance, 
$G$, decreases with the gate voltage, $V_{gate}$,  as shown by the measurement in Fig.\ref{fig3}(d), which was taken at 0.24 K. This p-type transistor behavior is characteristic of hole-dominated conduction. Once again, this follows from the fact that $\phi_p < \phi_n$.  An increase of $V_{gate}$ causes a downward band bending in the silicon section leading to a higher tunnel barrier for holes (Fig. \ref{fig3} (c)), and hence a lower conductance. 
Yet, due to a short-channel effect  (the nanowire diameter is about four times the channel length), the gate effect is largely screened by the metallic PtSi contact. As a result, the $G(V_{gate}) $ exhibits a moderate modulation over the accessible gate-voltage range. In particular, $G(V_{gate}) $ remains finite up to the highest gate voltage applied (no higher voltages could be achieved due to the onset of significant gate leakage).  A negative gate voltage produces an upward bending of the valence-band profile. Even for the largest $V_{gate}$, however, the valence-band edge remains well below the Fermi level of the contacts, such that no hole accumulation is induced in the silicon channel. This is consistent with the absence of Coulomb blockade behavior, which would be expected in concomitance with a gate-induced formation of a hole island in the silicon region  (see, e.g., Ref. \cite{Zwanenburg2009}).  
 
%DEVICE
%\section{Resonant tunneling through an isolated impurity}

\begin{figure}[h]
\includegraphics[width=8cm, keepaspectratio]{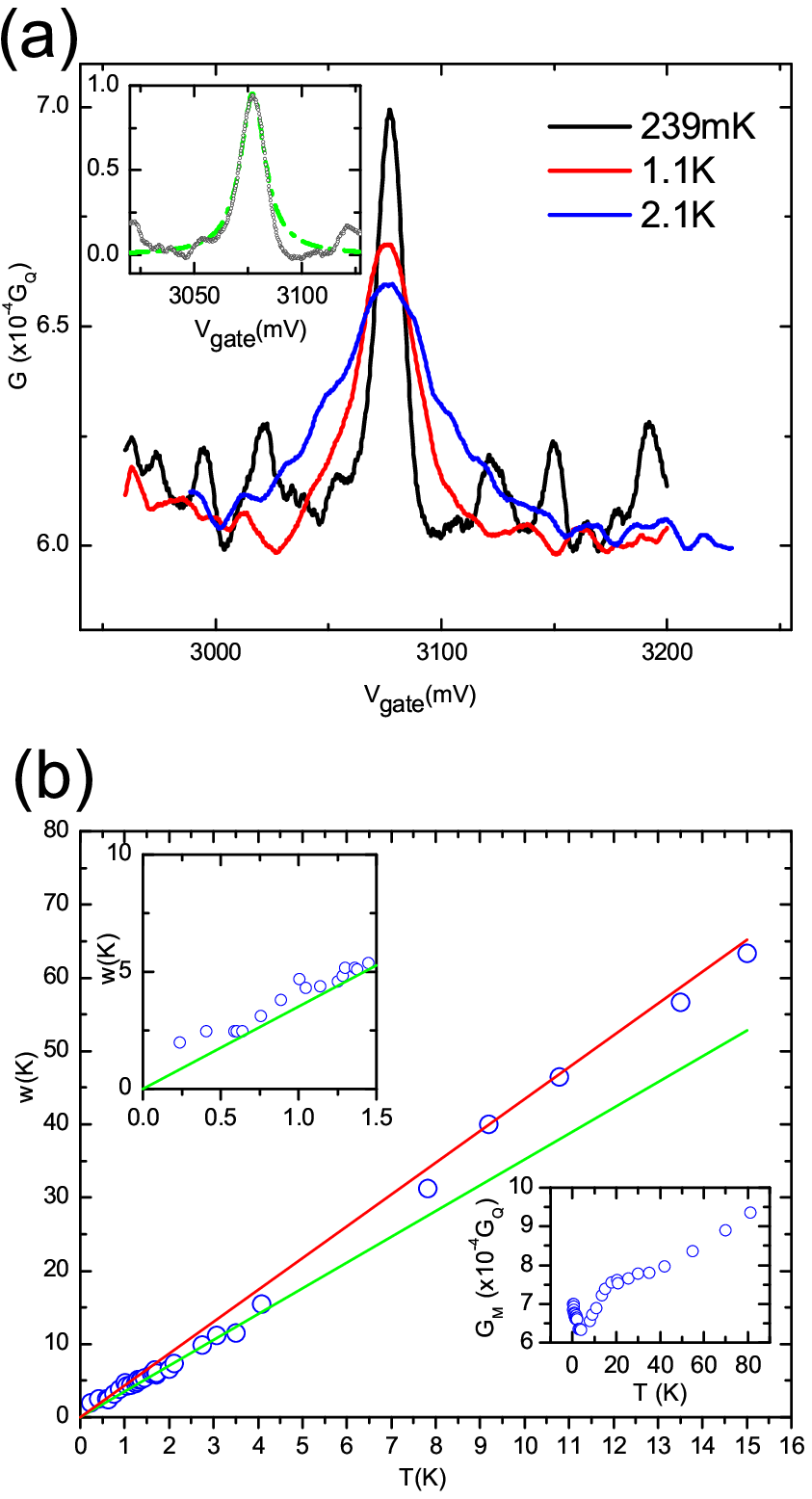}
\caption{(a)Thermal broadening for the resonant peak as in Fig. \ref{fig3}(d). As the temperature is lovered the resonance goes in the quantum regime and the peak becomes narrower and higher. The small shift in the gate voltage position of the resonant peak as compared as to Fig. \ref{fig3}(d) is due to offset charges induced by a large gate voltage sweep.
Inset: Conductance peak fitted by a Lorentzian (green dotted line) after substraction of the background current.   
The peak is fitted with the function $G(E)\propto\frac{e^{2}}{h}\frac{\Gamma_{S}\Gamma_{D}}{(E-E_{0})^{2}+w^{2}} $
where $E $ is the gate voltage-dependent energy of the resonant level and  $E_{0} $ the Fermi level in the leads,  to give an intrinsic  width, $w$, of the level of about $170\mu eV $ at $230$mK.
(b)Temperature dependence of $w $ at $V_{sd}=0$. The red and green solid lines are the expected slopes $3.52 k_B $ and $4.35 k_B  $ for the quantum and classical transport regimes respectively.  Upper left inset: close up view of the main panel at low temperatures where $w $ becomes temperature independent below $600mK $. Lower right inset: Conductance peak height plotted versus temperature.
}
\label{fig4}
\end{figure}

Interestingly, a sharp conductance peak  is observed at a positive gate voltage close to 3 V, superimposed on the slowly varying background conductance. As shown in the inset to Fig. \ref{fig4}(a),  following the subtraction of this background, the conductance peak can be fitted very well to a Lorentzian function \cite{Foxman1994,Beenakker1991}, revealing an underlying resonant-tunneling transport channel. The fitted peak width $w $ gives a measure of the tunnel coupling between the resonant state and the leads.
The temperature dependence of the observed conductance resonance is shown in Fig. \ref{fig4}(a). As expected for resonant-tunneling, the conductance peak gets smaller and wider upon increasing temperature from 0.24 to a $\sim 5$ K. Above $T \sim 5 K$, the peak height increases again as shown in the lower inset of Fig.  \ref{fig4}(b). On the contrary, $w$,  exhibits a monotonic temperature dependence as shown in Fig. \ref{fig4}(b)(main panel). To a closer look, however, $w$ is roughly constant below 0.6 K (see upper inset to Fig. \ref{fig4}(b)). In this low-temperature regime, the peak width is dominated by the life-time broadening of the resonant state due its tunnel coupling to the source and drain leads.  Precisely, $w \approx \hbar\Gamma = \hbar (\Gamma_{S} + \Gamma_{D}) $, where $\Gamma_{S} $ and  $\Gamma_{D}$ are the tunnel rates to the source and drain contacts, respectively.  Between 0.6 and 4 K, $w$ increases linearly with temperature according to the expectation for tunneling through a single discrete resonant level, i.e. $w = 3.52K_{B}T$ ((Fig. \ref{fig4}(b), green line). This linewidth  matches exactly the thermal broadening of the Fermi distribution function in the source and drain leads. A linear temperature dependence is observed also above 8 K, yet with a larger slope corresponding to the expectation for tunneling through an ensemble of closely spaced levels, i.e.  $w = 4.35 K_{B}T$ (Fig. \ref{fig4}(b), red curve). Therefore, the crossover temperature $T^{\star} \approx 6$ K identifies the transition from quantum (single level) to classical (multiple levels) regime \cite{DeFranceschi2003}. This finding suggests that the observed conductance peak arises from resonant tunneling through a quantum dot with characteristic mean-level spacing $\delta E \approx k_B T^{\star} \approx 0.5 meV$. 

\begin{figure}[h]
\begin{center}
\includegraphics[width=16cm,keepaspectratio]{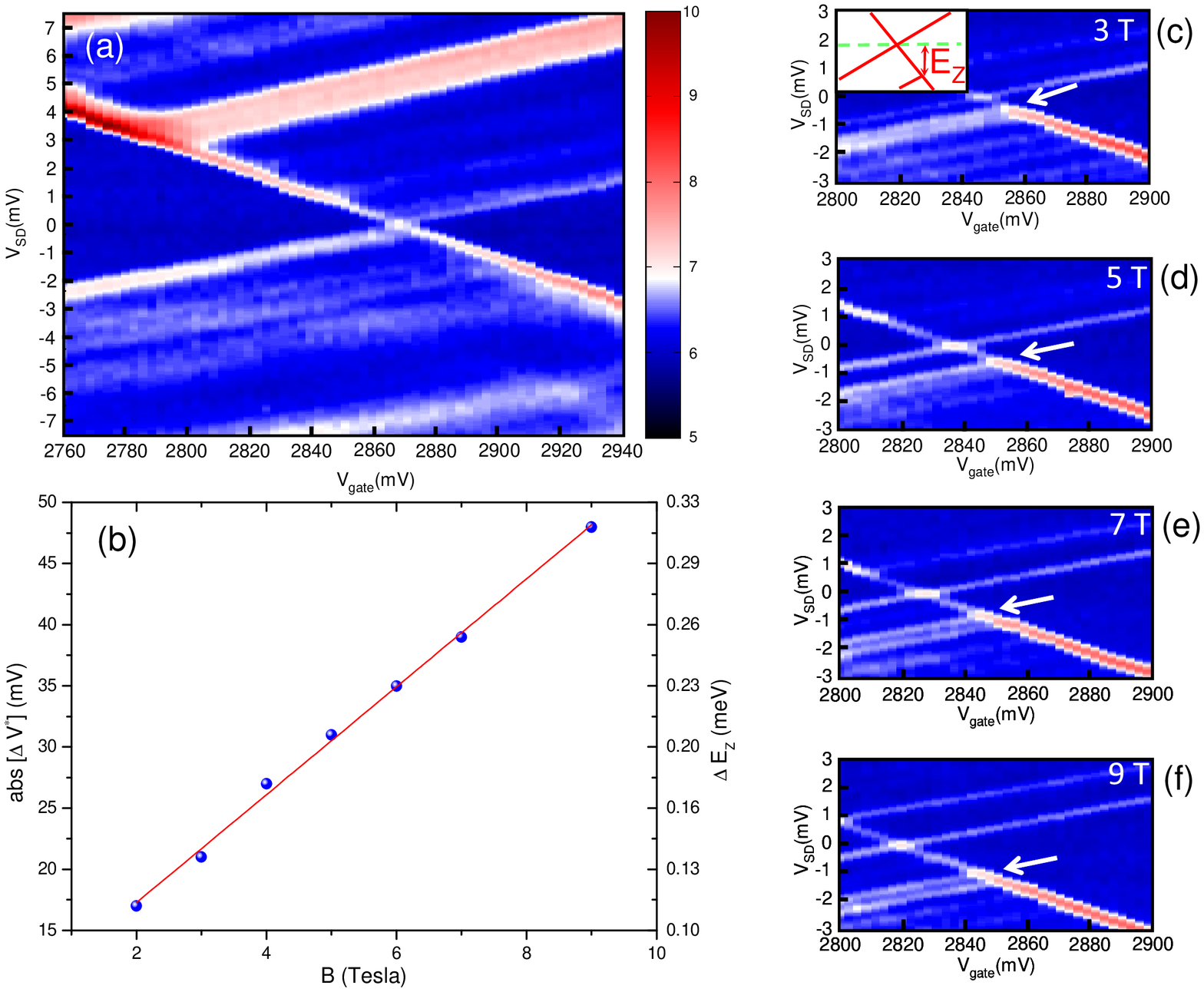}
\caption{  a) Differential conductance in units of  $10^{-4} e^{2}/h $ measured by varying the bias potential and the top gate voltage at magnetic field $B =0$. Lines of enhanced conductance due to the alignment of the resonant level to the Fermi energies in the leads, mark the separation between even and odd electron occupancies of the nanocluster. Using a conversion factor  determined by the slope of the differential conductance lines it is possible to convert in energy the positions in gate voltage ($V^{\star}$) of the resonant peak at $V_{SD}=0 $. The difference  $|\Delta V^{\star}|=|V^{\star}(B)-V^{\star}(B=0)| $ is then converted into energy  (right vertical scale, panel b) and a linear fit to the data (red line) gives a g-factor equal to  $2.0\pm 0.1$ c)-f) Same plot as in a) for B = 3, 5, 7, and 9 T, respectively. The white arrows highlight the evolution of the excited-spin-state resonance. The inset is c) shows how the Zeeman energy is related to the position of the excited-spin-state resonance.}
\label{fig5}
\end{center}
\end{figure}

To further support this conjecture we present in Fig. \ref{fig5}(a) a measurement of $dI_{sd}/dV_{sd}$ as a function of ($V_{gate}$,$V_{sd}$). The color plot (stability diagram) exhibits an X-shaped pattern which is typical for Coulomb-blockaded transport in quantum-dot systems \cite{Hanson2007,Kuemmeth2008}. The crossing point, which corresponds to the conductance peak in the linear regime, represents the boundary between two consecutive Coulomb diamonds. In each diamond, transport through the quantum dot is blocked and the quantum dot hosts a well defined, integer charge state.  Additional multiple $dI_{sd}/dV_{sd}$ lines parallel to the diamond edges can be seen in Fig. \ref{fig5}(a). Such types of lines are typical signatures of tunneling via the excited states of the quantum dot.  Their positions relative to the diamond edges are set by the excitation energies of the corresponding quantum-dot states.  The irregular spacing between the lines implies that the quantum-dot levels are not equally distributed in energy. Yet we can infer a characteristic mean-level spacing of the order of 1 meV, which is consistent with our earlier estimate based on the temperature dependence of the linear-conductance resonance. 

The next step is to understand what kind of quantum dot could be responsible for the observed Coulomb-blockade features. Based on the gate dependence of Fig. \ref{fig3}(d) we already concluded that the silicon channel is fully depleted throughout the entire gate-voltage range. This allows us to rule out the possibility that the quantum dot is formed by confinement in the short silicon section. Thus we are left with the sole option of a Pt-based metallic nanocluster. 
Assuming that electrons in the nanocluster can be described to a good approximation as non interacting quasi-particles in spin degenerate levels \cite{Delft2001}, the mean-level spacing can be derived from the cluster diameter, $d$, using the relation $\delta E \sim  2\pi^{2}\hbar^{2}/m k_{F} V $, where $m$,$k_F $ and $V $ are the electron mass, metal Fermi wavevector and nanoparticle volume, respectively. From a level spacing of $\sim$1 meV, using  $k_{F}\sim 3\cdot 10^{10} $m$^{-1} $ , we find $d \sim $ 4 nm. The data shown in Fig. \ref{fig5}(a) was focused on a narrow range around the observed tunnel resonance. From a similar measurement on a much larger ($V_g,V_{sd}$) range (see Supplemental Material) we estimate a charging energy $U \sim 50 meV$, which is consistent with a cluster diameter of a few nm.  

We now consider the effect of a magnetic field, $B$, on the observed resonant level. Figure \ref{fig5} (c)-(f) shows a set of stability diagrams measured at \textit{B} = 3, 5, 7, and 9 T (in all these measurements B was applied perpendicularly to the substrate).  We first note that by increasing \textit{B} the X-shaped structure shifts progressively towards less positive $V_{gate}$ values. 
This behavior can be ascribed to the Zeeman effect. To show that, let us label the gate-voltage position of the charge degeneracy point (i.e. the position of the resonance in the linear conductance) as $V^{\star}$; then let us start with the hypothesis that the nanocluster has an even number of electrons and a spin $S=0$ for $V_{gate}< V^{\star}$, and an odd number of electrons and S=1/2 for $V_{gate}> V^{\star}$. Under this hypothesis, an applied magnetic field should result in a negative shift of $V^{\star}$ proportional to the Zeeman energy shift $\Delta E_{Z}$ of the spin-1/2 ground state.

Figure \ref{fig5}(b) shows the gate-voltage shift, $\Delta V^{\star}$ as a function of $B$. The dashed line is a linear fit to $\Delta V^{\star} = \alpha \Delta E_Z = - \alpha g \mu_B B/2$, where $\alpha $ is a lever-arm parameter given by the ratio between the gate capacitance and the total capacitance of the cluster ~\footnote{$\alpha $ is a conversion factor which relates the potential on the gate to the potential in the nanocluster}, $\mu_B$ is the Bohr magneton, and $g$ is the electron g-factor, used as fitting parameter. The linear fit yields $g=2.0\pm 0.1$.

While the charge-degeneracy point undergoes a $B$-induced leftward shift,   the edges of the Coulomb diamond on its left are expected to split due to the removal of spin degeneracy. This effect is clearly seen on the lower diamond edge as highlithed by the white arrows in Figs. 5(c)-(f). The Zeeman energy can be directly measured from the position of the line associated to the excited spin state, as illustrated in the inset of Fig. \ref{fig5}(c). With this procedure we find $g =1.95 \pm 0.1$, which is consistent with our previous estimate.

The fact that the measured g-factor coincides (within the experimental uncertainty) with the bare-electron value allows us to make some considerations about the nature of the quantum dot. To begin with, this finding constitutes further (indirect) evidence that the observed resonant channel cannot be associated with an isolated Pt impurity. In fact, valence electrons in Pt impurities are known to have anisotropic g-factors generally different from the bare-electron value \cite{Woodbury1962,Anderson1991,Anderson1992}. This is because platinum is a heavy element with strong spin-orbit coupling. For the same reason, significant deviations of the g-factor from the bare-electron value have been observed in platinum metal structures \cite{Hjelm1991} (and refs. therein), including Pt nanoclusters \cite{Liu2006a} (we are aware of only one work \cite{Gordon} whose results do not agree with this trend). Therefore, our g-factor measurement provides an experimental indication against the hypothesis of a cluster consisting of pure platinum. The hypothesis of PtSi cluster, as suggested by our ab initio calculations, appears more plausible. With the aid of adequate numerical tools, it would be interesting to perform a calculation of the g-tensor in PtSi. Intuitively, one may indeed expect the g-factor to approach the bare electron value as a result of the considerable silicon content in the cluster. This tendency should be further reinforced by surface effects associated with the leakage of the electron wave-functions of surface atoms into the silicon host matrix \cite{Leeuwen1994,Buttet1982,Hai2010}.

In summary, through atomistic simulations and electronic transport measurements we have provided evidence of platinum clustering in silicon devices employing PtSi contacts. Our experiment on a short-channel PtSi/Si/PtSi SB transistor revealed the emergence at low temperature of a single-electron tunneling channel. This transport channel, which causes current resonances in the full-depletion (i.e. off) regime, is ascribed to a Pt-based metallic cluster embedded in the silicon section. Our atomistic simulations suggest the cluster to be most likely composed of PtSi. The cluster has discrete electronic levels with a characteristic energy spacing of 0.5 meV, corresponding to a diameter of a few nm, which is comparable to the characteristic length scales of the device. In addition,  the cluster has a charging energy of ~50 meV, which explains why single-electron effects survive up to relatively high temperatures. 
In the perspective of ultrascaled transistor devices, our study shows that the possible formation of PtSi clusters during the silicidation process can have important consequences on device performances. This issue can lead to significant device variability undermining the gain from using metal silicides and doping-free devices.  Hence, we expect our results to have an impact in the engineering of a wide class of emerging electronic devices, including fully-depleted nano-transistors and ultra-fast PtSi Schottky barrier photodetectors.

%CONCLUSIONS
\acknowledgments
This work was supported by the  Agence Nationale de la Recherche and by the EU through the ERC Starting Grant HybridNano . The authors would like to thank Laurent Cagnon Stephane Auffret and Xavier Waintal for technical support ans useful discussions. 

%\bibliographystyle{achemso}
%\bibliography{bibliography_nanowires}

\bibliography{bibliography-Pt}

\end{document}